\documentclass[mathleft
]{an}
\usepackage{graphicx}
\usepackage{times}
\begin{document}

\Pagespan{1}{}
\Yearpublication{}%
\Yearsubmission{2008}%
\Month{}%
\Volume{}%
\Issue{}%
\pretolerance=10000

\title{Radio Bimodality: Spin, Accretion Mode, or Both?}

\author{M. Sikora\inst{1}\fnmsep\thanks{Corresponding author:
  \email{sikora@camk.edu.pl}\newline}
}
\titlerunning{Radio Bimodality}
\authorrunning{M. Sikora}
\institute{
Nicolaus Copernicus Astronomical Center, ul. Bartycka 18, 00-716 Warsaw, 
Poland}

\received{13 Sep 2008}
\accepted{???}
\publonline{later}

\keywords{galaxies: nuclei -- galaxies: jets -- quasars: general}

\abstract{
A new scenario is suggested to explain a large diversity of the AGN radio
properties and their dependence on the galaxy morphology.
The scenario is based on the assumption that the growth of supermassive BHs 
is dominated by the accretion only during the quasar (high accretion rate)
phase, otherwise -
by mergers with less massive black holes. 
Following that, BHs are expected to spin much faster in  
giant ellipticals  than in disk galaxies.
Within the frame of the spin paradigm this explains the observed relation
of the radio-dichotomy with the galaxy morphology. 
Various theoretical and observational aspects of such a dichotomy
are discussed. In particular, the issue of the intermittency and suppression of 
a  jet production at high accretion rates is addressed and a scenario 
for production of powerful, extended radio sources is drafted.
}

\maketitle

\section{Introduction}
Diversity  of radio sources associated with AGNs is 
mainly  manifested by a very broad range of 
radio luminosities $L_{\mathrm{R}}$ and morphology. The most luminous are radio 
sources
powered by AGNs hosted by giant ellipticals (Xu, Livio \& Baum 1999). 
In terms of the radio-to-optical luminosity ratio,  
$R \propto L_{\mathrm{R}}/L_{\mathrm{O}}$, they
are by 2-3 orders radio louder than the most radio luminous AGNs hosted by
disk galaxies. This difference holds for a full range of 
the Eddington-ratio parameter, 
$\lambda \equiv L_{\mathrm{bol}}/L_{\mathrm{EDD}}$.
Despite such a big difference, maximal radio luminosities of
both populations
show similar dependence on $\lambda$. Those increase with $\lambda$ but
slower than linearly, which implies the opposite dependence of the
radio-loudness on $\lambda$, i.e.
$R$ increases with decreasing $\lambda$ (Sikora, Stawarz,\& Lasota 2007 
[SSL07]). While the dependence on $\lambda$ is most likely related to
the accretion rate, there must be an additional parameter which has the effect
on the jet production efficiency and is correlated with the galaxy 
morphology. SSL07 postulated that it is the black hole spin.
However for this, the two basic questions must be answered:
(1) how the cosmological evolution could lead to a black hole spin 
distribution weighted to much larger spin values in giant elliptical galaxies
than in disk galaxies; (2) why most quasars and other high accretion rate AGNs
are radio-quiet even if hosting fast spinning black holes.
These issues are the topics of sections \S2 and \S3, respectively,
and the work is summarized  in \S4.
 
\section{Spin vs. galaxy morphology}
In order to have the galaxy morphology-related spin bimodality,
SSL07 and Volonteri, Sikora \& Lasota (2007) 
proposed  the following scenario. They assumed that 
the growth of the  AGN black holes  is dominated 
by accretion: in quasars --- by very massive events triggered by major mergers
of gas rich galaxies, in AGNs hosted by the disk 
galaxies --- by multi-accretion events with random
angular momentum orientation and very small mass portions. They must be 
smaller than  
%
\begin{equation} m_{\mathrm{align}} \simeq 
\xi a \sqrt{r_{\mathrm{g}}\over r_{\mathrm{w}}} M_{\mathrm{BH}} \sim 
10^{-3} {(\xi/3) (a/0.1) \over 
\sqrt{r_{\mathrm{w}}/10^3r_{\mathrm{g}}}}M_{\mathrm{BH}} 
 \, , \end{equation}
%
which is the minimal mass of the accretion event leading to the alignment of 
the black hole spin with the angular momentum of the outer portions of 
an accretion  disk, $r_{\mathrm{g}} = \mathrm{G} M_{\mathrm{BH}}/c^2$ is 
the gravitational radius, $r_{\mathrm{w}}$ is the warp radius, 
$a \equiv J_{\mathrm{BH}}/J_{\mathrm{BH,max}} =
c J_{\mathrm{BH}}/\mathrm{G} M_{\mathrm{BH}}^2$ is the dimensionless  BH spin,
$J_{\mathrm{BH}}$ is the BH angular momentum, $M_{\mathrm{BH}}$ is the BH mass,
and $\xi=\nu_2/\nu_1$  where $\nu_1$ and $\nu_2$
are viscosities related to  the 'planar' and 'vertical' shear, respectively 
(Papaloizou \& Pringle 1983).
Only for the accretion events with masses $m_{\mathrm{acc}} < m_{\mathrm{align}}$
 the counter-rotating accretion disks 
avoid a flip to the co-rotating ones, and significantly smaller 
$m_{\mathrm{acc}}$ is required in order to balance the rate   
of counter-rotating and  co-rotating events
-- the condition needed to prevent growing BHs against gaining large spins
(King et al. 2005; Volonteri et al. 2007). 
However, such a condition is very difficult to satisfy because
accretion disks  in AGNs extend at least up to the broad emission line region 
and  their masses  are larger than $m_{\mathrm{align}}$ (but $\ll M_{\mathrm{BH}}$).
Furthermore, it is observed in some Seyfert galaxies that kiloparsec scale jets
are often bent (Gallimore et al. 2006). If these jet bends are caused by 
the BH-disk alignement process, this also implies 
$m_{\mathrm{acc}} \ge m_{\mathrm{align}}$.

No such difficulties arise if one  assumes that
the growth of BHs in the disk galaxies is dominated 
by capturing  BHs  
sinking to the center following mergers of a disk galaxy with dwarf satelite
galaxies (Hughes \& Blandford 2003; Kendall, Magorrian \& Pringle 2003). 
During such minor mergers, tidal forces from satelite galaxies are too weak
to induce the  inflow of the gas to the center. The satelite  
galaxies are stripped of the gas and enter the center 
from random directions. The process is finalized by a dry minor merger of two
very unequal-mass BHs and the sequence of such mergers cause the growth of 
the BH with a low spin. Cosmological evolution of BHs dominated by
the dry BH mergers was simulated by Volonteri (2007) and 
Berti \& Volonteri (2008).
They derived  much shallower distribution of BH spins than 
predicted by Hughes \& Blandford (2003). The reason is that 
time scale of dynamical friction  which drives the low mass satelite galaxies 
and their BHs toward the center of the more massive companion is very long 
and therefore major mergers are initially dominant. 
However, noting that up to masses  $2\times 10^6 M_{\odot}$ the BH growth
can be dominated by accretion of tidely 
disrupted stars (Milosavljevi\'c, Merritt \& Ho 2006), 
the spin distribution weighted to low values is very likely.

It is worth noting here that significant contribution
of BH mergers to the BH mass content of the local Universe is rejected by
most AGN-BH evolution models. This is primary because several
studies based on the So{\l}tan's type of argument (So{\l}tan 1982) have shown 
that in order to reconcile an amount of energy radiated by quasars 
per co-moving volume with a mass density of BHs in the local universe
requires
very efficient radiation even if assuming that the BH growth is totally
dominated by the accretion (see Wang et al. 2006 and refs. therein). However, 
the newest results obtained by  Shankar, Weinberg \& Miralda-Escud\'e (2007)
and Merloni \& Heinz (2008) suggest the   
lower energy requirements. Assuming that the BH growth is strongly
dominated by accretion, they found that radiation efficiency is consistent
with non-rotating BHs. But these same results can be interpreted 
in terms of the model where quasars radiate with efficiency corresponding 
with the fast rotating black holes provided  the BH growth in the pre-quasar
phase  
is significantly contributed by BH mergers (Cao \& Li 2008). Such an
interpretation is fully consistent with the scenario proposed above.

\section{Jet activity in the high accretion rate regime}
\subsection{Intermittency}
It has been proposed that there is an analogy between the jet production 
in X-ray binaries  and AGNs (Maccarone, Gallo \& Fender 2003).
Radio observations of the X-ray binaries indicate  
a suppression of the jet production at high
states  (Gallo, Fender \& Pooley 2003). However, there are also 
some 'sub-states' during which powerful jets are occasionally produced 
(Fender, Belloni \& Gallo 2004).  
It was suggested  that the low (3-10\%) fraction of the radio-loud quasars
is also due to the suppression of the jet 
production and  that  radio-loud quasars  are the 
products of the intermittent jet activity (Nipoti, Blundell \& Binney 2005;
K\"ording, Jester \& Fender 2006).  
Livio, Pringle \& King (2003) proposed the model according to which
intermittency is connected with stochastic transitions between two accretion 
modes, the standard one -- with angular momentum transmitted outwards by
viscous torques within a disk,  and the 'magnetical' one - with the
developed large scale poloidal magnetic fields and related MHD jets/winds.
SSL07 incorporated this idea into the spin paradigm scenario.
They assumed that  creation of narrow powerful jets requires both 
a large BH spin and an efficient collimation mechanism 
(Sol, Pelletier \& Asseo 1989; Begelman \& Li 1994) and postulated  
collimation of the central, Poynting flux dominated outflows
by heavier and slower MHD outflows generated intermittently in a disk. 

Intermittent production of jets  is imprinted in radio morphology of
some radiogalaxies. Best examples  are represented by galaxies
with double-double radio structures (see Saikia, Konar \& Kulkarni 2006 
and refs. therein). However, it is not clear whether
this intermittency is related to the changes of a disk accretion mode
taking place at the approximately constant accretion rate, or is 
reflecting a modulation of the accretion rate, e.g. by 
thermal-viscous instabilities (Siemiginowska \& Elvis 1997) or combination 
of the magnetorotational and gravitational instabilities 
in the outer portions of the accretion disk
(Menou \& Quataert 1997). Furthermore, existence of 
the double-double radio morphologies shows that the radio activity of jets
launched in the two different epochs overlap in time which means that 
the duty cycle is too large to interpret the radio-quiet quasars as those 
appearing between  phases of a jet production turn-off.  
One could  postulate that in most cases the duty cycle is 
much smaller, corresponding with the percentage of radio-loud quasars,
but this would imply the period of a total quasar phase exceeding
the Salpeter time already after the first cycle. 
More promising are models involving   much shorter time scales of 
intermittency. They were suggested
to explain an excess of the number density of  compact symmetric objects
over the number 
density predicted assuming that all these objects  represent
initial stages of expanding double radio sources. 
In the variant of a large duty-cycle the excess results from a delay of  
the expansion (Reynolds \& Begelman 1997), in the variant
of a small duty cycle the excess is explained by postulating  that most of 
double radio sources die at young ages (see Kunert-Bajraszewska \& Marecki 
2007 and refs. therein). Only in 
the latter case the intermittency can be reconciled with the very small 
percentage of radio-loud quasars.  Furthermore, detection of core-jet
structures in all lobe dominated radio-quasar indicates that there are no
turn-offs of a jet production in quasars with large scale radio structures
(Hough et al. 2002). Hence, these exceptional objects, constituting about 
3 \% of all quasars (de Vries, Becker \& White 2006; Lu et al. 2007),
must follow  a different evolutionary scenario than others.
It is suggested below.

\subsection{Quasars with powerful large scale jets}
Noting that it is much easier to develop  large scale magnetic fields 
in geometrically thick disks than in geometrically thin disks 
(Livio, Ogilvie \& Pringle 1999),
I assume that the  MHD winds/jets which collimate the central outflows 
are generated only  if  innermost
portions of the accretion disk or at least their surface layers are inflated.
The inflation can be induced intermittently by thermal instabilities 
(Janiuk, Czerny \& Siemiginowska 2002), or supported permanently   
by energy transfer from the fast rotating BH to the disk.
The latter  is expected to work efficiently in objects with BH spins 
$a>a_{\mathrm{eq}}$,
where $a_{\mathrm{eq}}$  is the maximum spin reachable via  the accretion  
process. Specifically: without any coupling and radiation $a_{\mathrm{eq}}=1$ 
(Bardeen 1970); with 
the radiation drag $a_{eq} \simeq 0.998$ (Thorne 1974); with the magnetic
coupling computed for MHD accretion disks with a moderate geometrical thickness
$a_{\mathrm{eq}} \simeq 0.93$ (Gammie, Shapiro \& McKinney 2004); and  smaller 
if counting the extraction of the BH energy and angular momentum via 
the Blandford-Znajek mechanism operating in an open magnetic field
configuration  (Moderski, Sikora \& Lasota 1998). 

Let consider now formation of quasars by two types of major mergers,
one involving  two gas-rich disk galaxies, and one where the disk galaxy 
is merged with a giant elliptical.  
In both cases the merger of two BHs presumably proceeds within a dense,
massive gas disk formed in the center of two colliding galaxies.
Such a  disk helps to drive the hardening of the BH binary on scales where
stellar dynamical friction is already inefficient and gravitational radiation
is still inefficient (Begelman, Blandford \& Rees 1980; 
Dotti, Colpi \& Haardt 2006; but see Cuadra et al. 2008).
The disk helps also to
avoid the ejection of the merged BHs from the center of the merging galaxies, 
by inducing the alignment of the  BH spins   
with the orbital spin prior to their coalescence, via the mechanism discussed 
at the beginning of \S2 (Bogdanovi\'c, Reynolds \& Miller 2007).
With such a configuration, a merger of two BHs with low initial spins and
mass ratio between 1:4 and 1:1 leads to formation of the BH
with the spin enclosed in the range between $\sim 0.5$ and $\sim 0.7$ 
(Rezzolla et al. 2007). Then, due to 
the accretion from the disk, BHs are spun-up to $0.8 < a_{\mathrm{eq}} < 0.9$. 
However, if one of the merging galaxies  is elliptical, 
hence already hosting the high spin BH,
a BH merger remnant will emerge with a spin $a \sim 0.96 > a_{\mathrm{eq}}$.
Following that we postulate that radio-loud quasars are the '2nd generation
quasars', triggered by collision of two galaxies with one already being 
the product of the previous major merger event.   This hypothesis is
consistent with recent studies of galaxy morphologies of radio-loud quasars
vs. radio-quiet quasars (Wolf \& Sheinis 2008). They  show that 
all radio-loud quasars reside in giant ellipticals while 
morphologies of galaxies hosting PG quasars are consistent with ongoing mergers 
of two gas-rich galaxies.

\subsection{Radio-intermediate quasars}
A large fraction of radio-loud quasars is compact (de Vries et al. 2006;
Lu et al. 2007). 
Some of them are as radio-luminous as extended double radio sources. They
are presumably connected with the intermittently launched jets and are 
represented
by compact symmetric objects. But most
of radio detected quasars have intermediate radio luminosities and it is not 
clear whether their radio activity is powered by jets. 
Their radio emission may originate, e.g. in shocks formed by the collision of
the uncollimated central outflows with the interstellar medium. 
This suggestion is supported by observations
of broad-absorption-line (BAL) quasars. These objects are preferentially found 
in intermediate-radio quasars with core dominated radio structures
(Liu et al. 2008). If assuming that their BAL systems 
result from the loading of the unconfined  central outflows by 
the cold filaments in the broad emission line region, 
the lack of BALs in radio-loud quasars 
with extended structures can be explained by the fact that in such quasars
central outflows
form well collimmated jets which are spatially separated from the broad 
emission line region.
In this picture, intermediate radio quasars may represent the tail of the 
radio-quiet population of quasars (Zamfir, Sulentic \& Marziani 2008)
and the lack of radio emission in most quasars can be explained assuming
that in most cases central outflows are effectively slowed down by the mass 
entrainment and do not form the terminal shocks.

\subsection{Radiogalaxies}
High accretion rate radio galaxies with extended FRII type double 
radio structures are represented by so-called broad-line-radio-galaxies (BLRG),
with optical morphology of giant ellipticals. They are as radio-luminous 
as radio loudest quasars,
and on the radio-loudness vs. Eddington-ratio plot form smooth extension 
of radio-loud quasars toward  lower $\lambda$'s (SSL07).  
They, similarly to radio-loud quasars, constitute only 
a small fraction of AGNs hosted by giant ellipticals. Most are radio 
intermediate or weak (Kauffmann, Heckman \& Best 2008) and the cause can  be 
the same as proposed for quasars, i.e. the lack of collimation of the central 
outflow.

\subsection{Seyfert galaxies}
According to the scenario sketched in \S2, AGNs in disk galaxies,
 are not related to any specific merger process but are occasionally triggered 
by gravitational instabilities secularly developed  in the galactic disks fed
by the gas stripped from the satelite galaxies and by the cold streams from 
the halo (Johansson, Naab \& Burkert 2008; Bournaud, Jog \& Combes 2007).
As observations of Seyfert galaxies indicate most of these AGNs 
radiate at intermediate rates which combined with their lifetimes 
give the mass increment not enough large to spin-up BHs significantly.
However there is a category of Seyfert galaxies, called NLS1 
(narrow-line Sefert1) which accrete at very high rates and,
therefore, like  quasars may host fast spinning BHs.
Interestingly, as recent surveys show, some of  NLS1s are 
radio-loud (Yuan et al. 2008). The radio-loud fraction of NLS1 is  smaller
than the radio-loud fraction of quasars which may indicate a lower duty 
cycle of the intermittent jet production in NLS1 or that not in all of them
the high accretion lasts sufficiently long to significantly spin-up BHs. 

\section{Summary}
Radio studies of AGNs at low accretion rates indicate that AGNs hosted 
by elliptical galaxies are on average by 2-3 orders radio louder
than AGNs hosted by disk galaxies (see SSL07 and refs.therein).
This can be interpreted  in terms of the spin paradigm assuming
that BHs in the elliptical galaxies gain during their evolution much larger 
spins than BHs in the disk galaxies. 
SSL07 and Volonteri et al. (2007) assumed that growth of BHs in all galaxies
is dominated by accretion and showed that low spin BHs in disk galaxies
may be due to their accretion history composed of many 
small mass accretion events with random angular momenta.
However noting that masses of 
accretion disks in Seyfert galaxies are larger than the upper mass limit
per the accretion event (see \S2) and 
motivated by new results indicating   lower quasar radiative ouput  
per BH mass than previously claimed, it is postulated  in this work that 
the BH growth 
in the disk galaxies is dominated by BH mergers, 
with  BHs captured from satelite galaxies.

This work also suggests a new scenario to explain the production of 
large scale jets powering extended double radio sources  in quasars and BLRGs. 
The scenario is based on the assumption that initial
collimation of the centrally produced jets can be provided by MHD
jet/wind from the inflated innermost portions of the accretion disk
and that such inflation may persist for very long time if supported by the 
transfer of the energy from the fast rotating BH to the disk.
This  transfer is expected to be  efficient 
if the BH is spinning  with $a>a_{\mathrm{eq}}$, and it is 
argued that such a spin  is available
in the 2'nd generation of quasars, i.e. following a major merger of a
giant elliptical (a product of the past quasar event) with a gas-rich disk 
galaxy.

\acknowledgements
I am grateful to Roya Mohayaee, Isaac Shlosman, Jean-Pierre Lasota, and Greg 
Madejski for inspiring discussions.
The work was partially supported by the University of Pierre and Marie Curie. 
Hospitality of IAP is acknowledged where most of the work was completed.

\newpage

\end{document}